\documentclass[
    ,final            
  ]
  {aipproc}

\layoutstyle{6x9}

\begin{document}

\title{Neutron structure function moments at leading twist}


\author{M. Osipenko}{
  address={Istituto Nazionale di Fisica Nucleare, Sezione di Genova, Genoa, Italy 16146}
}


\author{S. Simula}{
  address={Istituto Nazionale di Fisica Nucleare, Sezione Roma III, Roma, Italy 00146}
}

\author{S. Kulagin}{
  address={Institute for Nuclear Research of Russian Academy of Science, Moscow, Russia 117312}
}

\author{G. Ricco}{
  address={Universit\`a di Genova, Genoa, Italy 16146}
}

\author{CLAS Collaboration}{address={Jefferson Lab, Newport News, Virginia 23606}}

\begin{abstract}
The experimental data on $F_2$ structure functions of the proton and deuteron
were used to construct their moments. In particular, recent measurements performed
with CLAS detector at Jefferson Lab allowed to extend our knowledge of
structure functions in the large-$x$ region. The phenomenological analysis
of these experimental moments in terms of the Operator Product Expansion
permitted to separate the leading and higher twist contributions.
Applying nuclear corrections to extracted deuteron moments we
obtained the contribution of the neutron. Combining leading twist moments of
the neutron and proton we found $d/u$ ratio at $x\to 1$ approaching $0$,
although $1/5$ value could not be excluded.
The twist expansion analysis suggests that the contamination of higher twists
influences the extraction of the $d/u$ ratio at $x\to 1$ even at $Q^2$-scale
as large as 12 (GeV/c)$^2$.
\end{abstract}

\maketitle


\section{Data Analysis}
QCD through the Operator Product Expansion (OPE) allows
to relate measurable moments of nucleon structure functions to
the series of local operators, so called twists.
The Leading Twist (LT) (first term in the series) represents the asymptotic freedom
domain. This term is completely determined by 
perturbative calculations and Lattice simulations\footnote{Lattice simulations
up to now are limited to a few lower moments}.
Higher Twists (HT) (all further terms) describe the virtual photon
scattering off interacting partons.
The complexity of this interaction and therefore of corresponding QCD operators increases
with twist order. Calculations of these terms have been performed
only in a few cases.

\begin{figure}\label{fig:moms}
\includegraphics[bb=2cm 6cm 22cm 24cm, scale=0.4, height=.3\textheight]{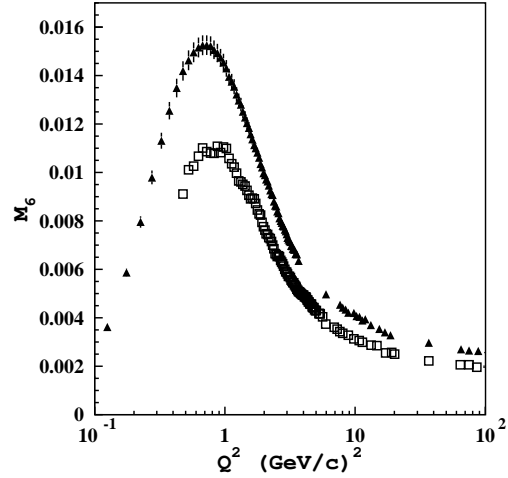}
\caption{Total experimental $n=6$ moments of the proton (full triangles) and
deuteron per nucleon (open squares) structure functions $F_2$.}
\end{figure}

The experimental data on the structure function moments were obtained
recently for the proton and deuteron in Refs.~\cite{osipenko_f2p} and \cite{osipenko_f2d},
respectively.
Moments were extracted from a combined analysis of the structure functions $F_2$
measured at CLAS and other world data on the inclusive lepton-nucleon scattering.
This was performed by integrating experimental data points independently
at each fixed $Q^2$ value. Hence, the obtained $Q^2$-evolution of the moments
is not affected by any model dependence and can be directly compared
to pQCD predictions. Example of measured moments is shown in Fig.~\ref{fig:moms}.
As one can see the proton and deuteron moments have similar $Q^2$-behavior,
but different absolute value. Indeed, in the QCD at the LT two moments
should have the same $Q^2$-evolution. This observation suggests that
the duality is valid for both targets and the contribution of nuclear
HTs\footnote{nuclear higher twists are mostly related to Final State Interactions
of the nucleon in the nucleus} is small in the covered $Q^2$-range.

\begin{figure}\label{fig:hts}
\includegraphics[bb=2cm 6cm 22cm 24cm, scale=0.4, height=.3\textheight]{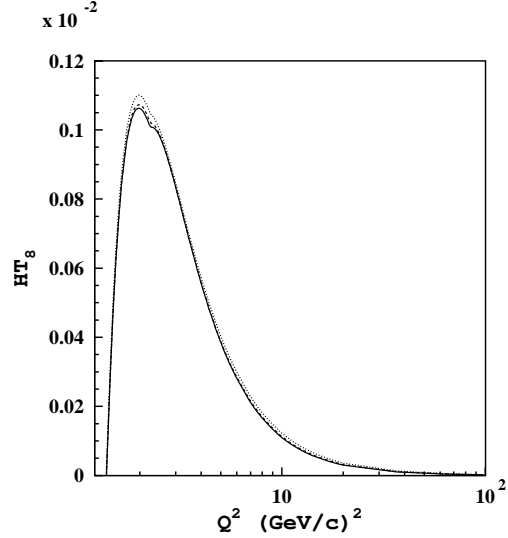}
\caption{Total higher twist contribution in $n=8$ moment obtained with the present
procedure for different number of terms in the OPE series (see Eq.~\ref{eq:ope}):
solid line - two HT terms,
dashed line - three HT terms,
dotted line - four HT terms.}
\end{figure}

We analyzed these experimentally extracted moments of the
proton and deuteron structure functions $F_2$~\cite{osipenko_f2p,osipenko_f2d}
to separate LT and HT terms. This was performed by fitting the data with
the following expression:
\begin{equation}\label{eq:ope}
M_n (Q^2)= \int_0^1 dx x^{n-2} F_2(x,Q^2) = LT_n(\alpha_S)
+ \sum_{\tau=4}^k a_n^\tau
\Biggl(\frac{\alpha_S(Q^2)}{\alpha_S(\mu^2)}\Biggr)^{\gamma_n^\tau}
\Biggl(\frac{\mu^2}{Q^2}\Biggr)^{\frac{\tau-2}{2}} ~,
\end{equation}
\noindent where $\alpha_S$ is the running coupling constant,
$\mu^2$ is an arbitrary scale (taken to be 10 (GeV/c)$^2$),
$a_n^\tau$ is the matrix element of corresponding QCD operators,
$\gamma_n^\tau$ is the anomalous dimension,
$\tau$ is the order of the twist and $k$ is the maximum HT order considered.
The number of HT terms ($k$) in the expansion is of course
arbitrary because we don't know at which $1/Q^2$ power
the series converges. This prevents an evaluation of each separate
HT term from the data. Instead, {\em the total contribution of HTs
can be extracted with good precision}. In Fig.~\ref{fig:hts}
one can see that taking two, three or four HT terms in Eq.~\ref{eq:ope}
does not change the total HT contribution. Moreover, this result
was expected because the total HT contribution represents
simply the difference between the data and calculated LT.

Extracted LT components of the proton and deuteron moments can be combined
now to obtain moments of the neutron structure function $F_2$.
In the Euclidean space of moments the convolution of the nuclear Impulse Approximation (IA)
transforms into a product of moments. This allows for a simple
extraction of neutron moments from the following algebraic relation:
\begin{equation}\label{eq:nucl_cor}
M_n^n(Q^2)=\frac{2M_n^D(Q^2)}{N_n^D}-M_n^p(Q^2) ~,
\end{equation}
\noindent where $M_n^p$, $M_n^n$ and $M_n^D$ are moments of the proton, neutron and deuteron,
respectively. $N_n^D$ is the moment of the nuclear momentum distribution $f^D$
i.e. the structure function of the deuteron composed of point-like nucleons
(see Ref.~\cite{osipenko_f2n} for details).
This nuclear structure function $f^D$ was obtained from the data on
the deuteron wave function.

\begin{figure}\label{fig:udr}
\includegraphics[bb=2cm 6cm 22cm 24cm, scale=0.4, height=.3\textheight]{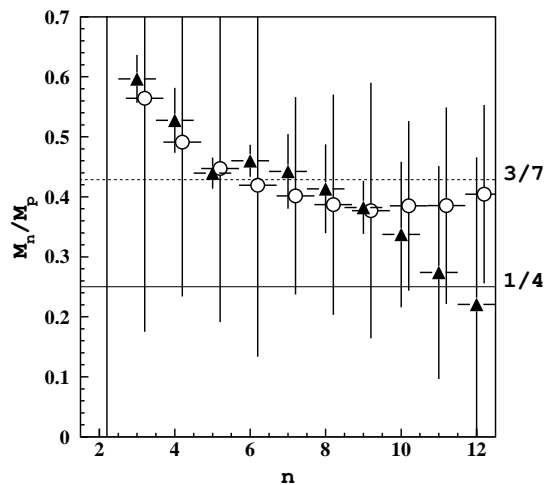}
\caption{Ratio of the neutron to proton moments as a function of $n$ at $Q^2=12$ (GeV/c)$^2$:
full triangles show the ratio at the leading twist obtained in Ref.~\cite{osipenko_f2n},
open circles represent the ratio of measured inelastic moments
including the higher twist contribution.}
\end{figure}

The data on proton and neutron moments can be used to study
the contribution of $u$ and $d$ quarks in the proton at $x\to 1$.
Assuming that $u$ and $d$ quark distributions in the proton
and neutron are the same\footnote{though the number of $u$ and $d$ quarks
is of course different} the $d/u$ ratio {\em at the leading twist accuracy} can be related to
the ratio of neutron to proton structure functions $F_2^n/F_2^p (x\to 1)$.
This ratio, in turn, is equal to the ratio of moments of these structure
functions $M_n^n/M_n^p (n\to \infty)$.
The latter equality requires only that $F_2^{p,n} (x\to 1) \to a^{p,n} (1-x)^{b^{p,n}}$,
which follows from the analyticity of the forward Compton amplitude in OPE.
Instead, the existence of a finite $F_2^n/F_2^p (x\to 1)$ limit guarantees
that exponents $b^p$ and $b^n$ are exactly the same.
We constructed these ratios of moments
and results are shown in Fig.~\ref{fig:udr}. As one can see our
data tend to the ``standard'' $1/4$ value used in most of parton distribution fits.
This value corresponds to the vanishing $d/u$ ratio. However, the precision
of our data does not allow to exclude $3/7$ value corresponding to $d/u$ ratio $1/5$.
In the Fig.~\ref{fig:udr} one also can see the impact of the HT contribution
on the ratio at $Q^2=12$ (GeV/c)$^2$. The previous analysis~\cite{Melnitchouk}
performed in the $x$-space showed that applying similar nuclear corrections
the ratio of structure functions $F_2^n/F_2^p(x\to 1)$ goes to $3/7$
at $Q^2=12$ (GeV/c)$^2$. No corrections on the possible HT contamination
has been applied in this article, assuming that at such large $Q^2$ they are
negligible. Therefore, if instead of LT part we take measured inelastic moments,
containing also HT terms, and construct the same ratios we should confirm
the result from Ref.~\cite{Melnitchouk}. Indeed, in the Fig.~\ref{fig:udr}
one can see that the ratio of moments including HTs tends to $3/7$ at largest $n$
($n=12$ corresponds to $x$ values about 0.75). This result could also be deduced
from the isospin independence of HTs observed in the Ref.~\cite{osipenko_f2n}.
\section{Conclusions}
We obtained the experimental data on moments of proton and deuteron structure function $F_2$.
In these data contributions of the leading and higher twists were separated via OPE analysis.
By combining the proton and deuteron moments and applying nuclear corrections
we extracted moments of the neutron structure function $F_2$.
The ratio of the neutron and proton moments at large $n$ is related to the ratio
of $d$ and $u$ quark contributions in the proton at large-$x$.
The obtained ratio is consistent with the asymptotic limit $d/u\to 0$ at $x\to 1$,
which originates from the dominance of soft, non-perturbative physics at large-$x$.
Nevertheless the alternative value of $1/5$, derived from helicity conservation
arguments, is not excluded.
The HT contamination to the ratio, if not subtracted as in the present analysis,
influences significantly the extracted $d/u$ ratio also at large $Q^2$-values.




\bibliographystyle{aipproc}   

\bibliography{Mikhail_Osipenko}

\IfFileExists{\jobname.bbl}{}
 {\typeout{}
  \typeout{******************************************}
  \typeout{** Please run "bibtex \jobname" to optain}
  \typeout{** the bibliography and then re-run LaTeX}
  \typeout{** twice to fix the references!}
  \typeout{******************************************}
  \typeout{}
 }

\end{document}